\documentclass[usenatbib]{mn2e}
\usepackage{graphicx}

\def\apj{\rm ApJ}
\def\apjl{\rm ApJL}

\def\aj{\rm AJ}
\def\mnras{\rm MNRAS}
\def\nat{\rm Nature}

\def\pasp{\rm PASP}

\def\aap{\rm AAP}
\def\araa{\rm ARA\&A}

\def\gax{\mathrel{\raise.3ex\hbox{$>$}\mkern-14mu\lower0.6ex\hbox{$\sim$}}}
\def\lax{\mathrel{\raise.3ex\hbox{$<$}\mkern-14mu\lower0.6ex\hbox{$\sim$}}}
\def\gtorder{\mathrel{\raise.3ex\hbox{$>$}\mkern-14mu
             \lower0.6ex\hbox{$\sim$}}}
\def\ltorder{\mathrel{\raise.3ex\hbox{$<$}\mkern-14mu
             \lower0.6ex\hbox{$\sim$}}}

\begin{document}

\title 
   [The Physics of Flash (Supernova) Spectroscopy]
   {The Physics of Flash (Supernova) Spectroscopy}

\author[C.~S. Kochanek]{ 
    C.~S. Kochanek$^{1,2}$
    \\
  $^{1}$ Department of Astronomy, The Ohio State University, 140 West 18th Avenue, Columbus OH 43210 \\
  $^{2}$ Center for Cosmology and AstroParticle Physics, The Ohio State University,
    191 W. Woodruff Avenue, Columbus OH 43210 \\
   }

\maketitle

\begin{abstract}
We examine flash spectroscopy of a circumstellar medium (CSM) ionized 
by the hard radiation pulse produced by the emerging shock of a supernova (SN).  We first find that
the rise and fall times of the H$\alpha$ emission constrains the location of the CSM 
with a peak at $t_{peak} \simeq R_* \sqrt{2/c v_s}$ 
for a star of radius $R_*$ and a shock velocity of $v_s$.  The dropping temperature
of the transient emission naturally reproduces the evolution of lines with 
different ionization energies.  Second, for red supergiants (RSGs), the shock break out
radiatively accelerates the CSM to produce broad, early-time line wings independent
of the Thomson optical depth of the CSM.  Finally, the CSM recombination rates in
binaries can be dominated by a dense, cool, wind collision interface like those seen in
Wolf-Rayet binaries rather than the individual stellar winds.  Combining these three
results, the flash spectroscopy observations of the normal Type~IIP iPTF13dqy (SN~2013fs) 
are naturally explained by an RSG with a normal, Thomson optically thin wind in a binary 
with a separation of $\sim 10^4 R_\odot$ without any need for a pre-SN eruption.
Similarly, the broad line wings seen for the Type~IIb iPTF13ast (SN~2013cu), whose
progenitors are generally yellow supergiants in binaries, are likely due to radiative 
acceleration of the CSM rather than a pre-existing, Wolf-Rayet-like wind.
\end{abstract}


\section{Introduction}
\label{sec:introduction}

Some massive stars appear to know that they are about to die.  They
manifest their impending death through outbursts that eject 
significant amounts of mass shortly before a supernova (SN), where
the outbursts are either seen directly (e.g., \citealt{Pastorello2007}, 
\citealt{Fraser2013}, \citealt{Mauerhan2013},
\citealt{Ofek2014}, \citealt{Ofek2016}) or inferred from
evidence for a massive circumstellar medium (CSM) once the
SN occurs (e.g., \citealt{Galyam2012}, \citealt{Smith2014b}).  Such systems are relatively rare, and
a range of theoretical models have been proposed to explain
the phenomenon (e.g., \citealt{Quataert2012}, \citealt{Shiode2014}, \citealt{Smith2014},
\citealt{Woosley2015}, \citealt{Fuller2017}). 

One means of probing the CSM is to use ``flash spectroscopy''
of the CSM shortly after a supernova is discovered (\citealt{Galyam2014},\citealt{Khazov2016}).
The burst of ionizing photons created by the SN shock breaking
out of the surface of the star flash ionizes the CSM and
the properties of any narrow emission lines in 
spectra of the SN taken before the CSM material can
either recombine or be swept up by the expanding shock
probes the density and composition of the CSM. \cite{Groh2014}
and \cite{Grafener2016} are able to match the early time
spectra using models of Wolf-Rayet stars, consistent with
a CSM photoionized by a radiation from a shock break out.
 
The most extensive flash spectroscopy observations are for
the normal Type~IIP iPTF13dqy (SN~2013fs, \citealt{Yaron2017}).
Based on the line fluxes and widths, \cite{Yaron2017} concluded that
the CSM density implied a far higher mass loss rate 
($\dot{M} \sim 10^{-3}M_\odot$/year) than is typical
of red supergiants.  Since the SN also shows no signs of 
strong CSM interactions (i.e., radio or X-ray emission) in its 
later phases, they also conclude that the dense CSM can 
only extend to $\ltorder 10^{15}$~cm.  \cite{Yaron2017}
argue that the progenitor must have experienced a 
short-lived, high-mass loss precursor before exploding
and that such precursors must be common.

As part of our search for failed SNe forming black holes
with the Large Binocular Telescope (LBT, \citealt{Gerke2015},
\citealt{Adams2016}), we also 
obtain high precision light curves of the progenitors
to any successful SNe in the target galaxies.  
We have monitored four Type~IIP progenitors
to date, finding no evidence for variability
down to levels close to the typical variability
of red supergiants (\citealt{Kochanek2017},
\citealt{Johnson2017}).  The absence of any 
visible outbursts or evidence of dust formation, which is essentially
inevitable at high mass loss rates ($\gtorder 10^{-4}M_\odot$/year,
see, e.g., \citealt{Kochanek2011}), 
strongly argues that outbursts from the progenitors
of Type~IIP SNe are in fact uncommon.  This is
further supported by the lack of X-ray emission from
normal Type~IIP SNe (e.g., \citealt{Dwarkadas2014})
although there are exceptions like SN~2013ej 
(\citealt{Chakraborti2016}). One counterargument by 
\cite{Morozova2017} and \cite{Morozova2018} is that 
a period of enhanced mass loss shortly before explosion
would help to explain the shapes of Type~IIP light curves.

This contradiction motivated an investigation into flash
spectroscopy with the goal of better understanding
iPTF13dqy.  We investigate three
issues, each of which has general applicability,
and then apply them to iPTF13dqy.  In \S2 we 
examine the time evolution of the line fluxes 
due to finite light travel times and recombination
as the SN cools and becomes less luminous.  We
find that the time evolution of the lower 
ionization energy lines (e.g., H$\alpha$) determines 
the location of the emitting material.  
In \S3 we show that the radiative acceleration of the
CSM by the shock break out radiation from an RSG
will produce broad line wings in early-time spectra
independent of the Thomson optical depth of the
CSM.  In \S4, we show that a radiatively cooling
boundary layer produced by the collision of the 
winds in a binary, as seen in some Wolf-Rayet binaries,
can produce a high density region that mimics a 
high mass loss rate from a single star.    

When we apply these general results to iPTF13dqy,
we come to the conclusion that emission from a
wind collision region provides a more natural
explanation of the observations than a pre-SN
eruption.  The time evolution of the H$\alpha$
emission requires material spanning a wide range
of radii that is also  detached from the stellar 
surface. The effects of radiative acceleration mean
that a Thomson optically thick CSM is no longer
needed.  The overall line luminosities still 
require denser material than a normal RSG wind,
but this is naturally supplied by a cooling wind
collision region.  We summarize both the general
and specific conclusions in \S5. Two technical
points are discussed in short appendices. 

\begin{figure}
\centering
\includegraphics[width=0.45\textwidth]{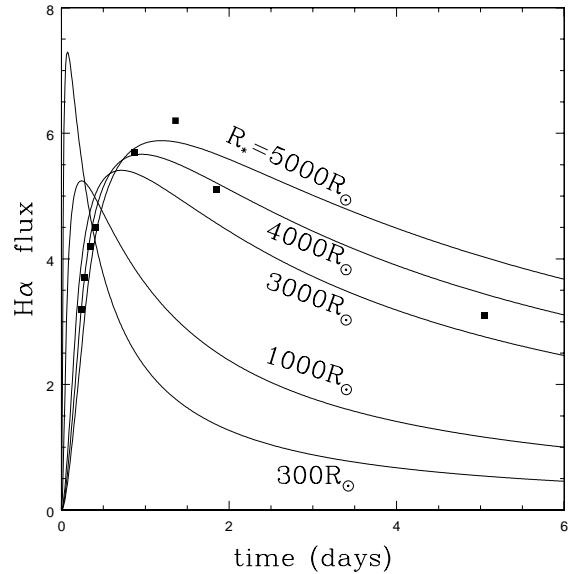}
\caption{ 
  Recombination light curves for a wind with an inner edge at a shock radius expanding
  at $v_{s0}=10^4$~km/s and an outer edge expanding at the speed of light, both starting
  from an initial radius of $R_{*}$.  This includes the effect of light travel times
  but ignores any shadowing effects (Equations~\protect\ref{eqn:delay1} and \protect\ref{eqn:delay1b}).  
  The points are the observed H$\alpha$ light curve from \protect\cite{Yaron2017} and 
  the curves are normalized to fit the equally weighted observations.
  The rise time requires a stellar radius of $R_{*} \simeq 4000R_\odot$,
  which is not physically reasonable for a red supergiant.
  }
\label{fig:lc}
\end{figure}

\begin{figure}
\centering
\includegraphics[width=0.45\textwidth]{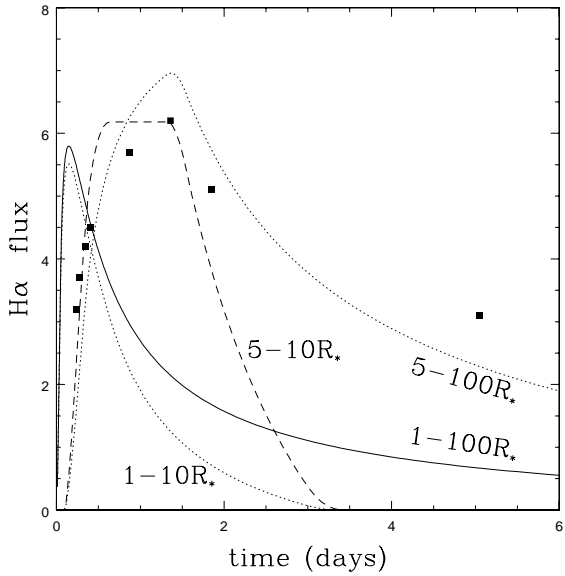}
\caption{ 
  Recombination light curves for a $R_*=821 R_\odot$ star with a dense wind lying between
  $1$-$100 R_*$ (solid).  The dotted line shows the effect of truncating the wind from
  the outside ($1$-$10R_*$) or the inside ($5$-$100 R_*$) and the dashed line shows
  the effect of making the wind a shell ($5$-$10R_*$).  An inner edge of $ 5 R_* = 4100 R_\odot$
  reproduces the rise in the $H\alpha$ flux well, as expected from Figure~\protect\ref{fig:lc}.
  The late time H$\alpha$ point requires an extended wind ($100 R_* \gtorder 6 \times 10^{15}$~cm)
  rather than a wind truncated on the scale of $10 R_* \simeq 6 \times 10^{14}$~cm 
  proposed by \protect\cite{Yaron2017}. The assumptions are otherwise the same as
  in Figure~\protect\ref{fig:lc}.  
  }
\label{fig:lc2}
\end{figure}

\begin{figure}
\centering
\includegraphics[width=0.45\textwidth]{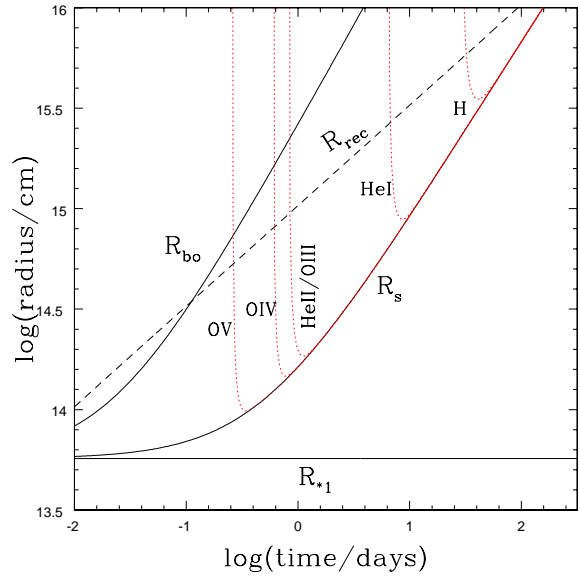}
\caption{ 
  A space-time diagram showing the evolution of radii relevant to interpreting
  flash spectroscopy.  A horizontal line shows the stellar radius $R_{*}=821R_\odot$
  chosen to match the $15M_\odot$ progenitor model of \protect\cite{Woosley2002}.  $R_{bo}(t)$ shows
  the expansion of the ionization wave created by the shock break out radiation
  pulse, and $R_s(t)$ shows the expansion of the shock. The diagonal line
  $R_{rec}(t)$ shows the radius where the recombination time equals the 
  elapsed time for the assumed wind properties of $\dot{M}=10^{-3.5}M_\odot$/year
  and $v_w=76$~km/s.  The curved dotted lines show, from left to right, 
  the radii $R_{pi}(t,E_\gamma)$ to which the current rate of production
  of ionizing photons above $E_\gamma$ can ionize OV, OIV, HeII/OIII and
  HI. The radii plunge very rapidly due to the exponential decline of 
  the ionizing fluxes and then turn and asymptotically track the shock
  radius. 
  }
\label{fig:spacet}
\end{figure}

\begin{figure}
\centering
\includegraphics[width=0.45\textwidth]{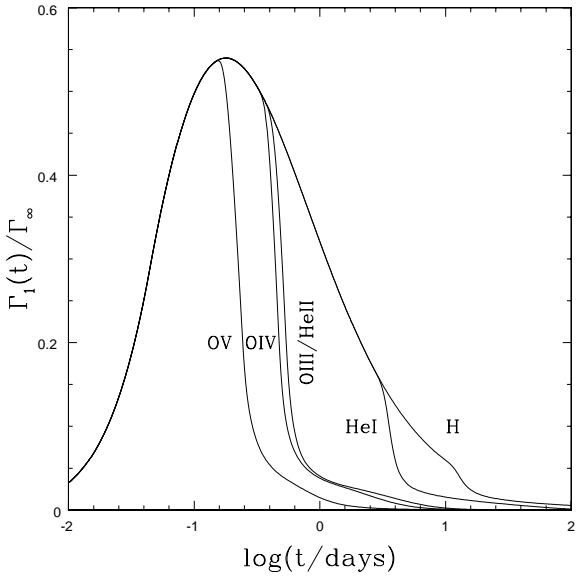}
\caption{ 
  Recombination rates to OV, OVI, OIII/HeII, HeI and H 
  relative to $\Gamma_\infty$ as a function of time.  The
  outer envelope is essentially the same as in 
  Figure~\protect\ref{fig:lc} except the shock is 
  decelerating to follow the self-similar solution.
  The recombination rates collapse when the numbers
  of ionizing photons become too small to maintain
  an ionized wind.   The model evolves more rapidly
  than iPTF13dqy because the fiducial model has 
  $R_*=821R_\odot$ while the data appear to require
  $R_*\simeq 4000 R_\odot$.
  These are for the full numerical
  calculation including a slowing shock front, recombination and
  light propagation effects.
  }
\label{fig:recomb1}
\end{figure}

\begin{figure}
\centering
\includegraphics[width=0.45\textwidth]{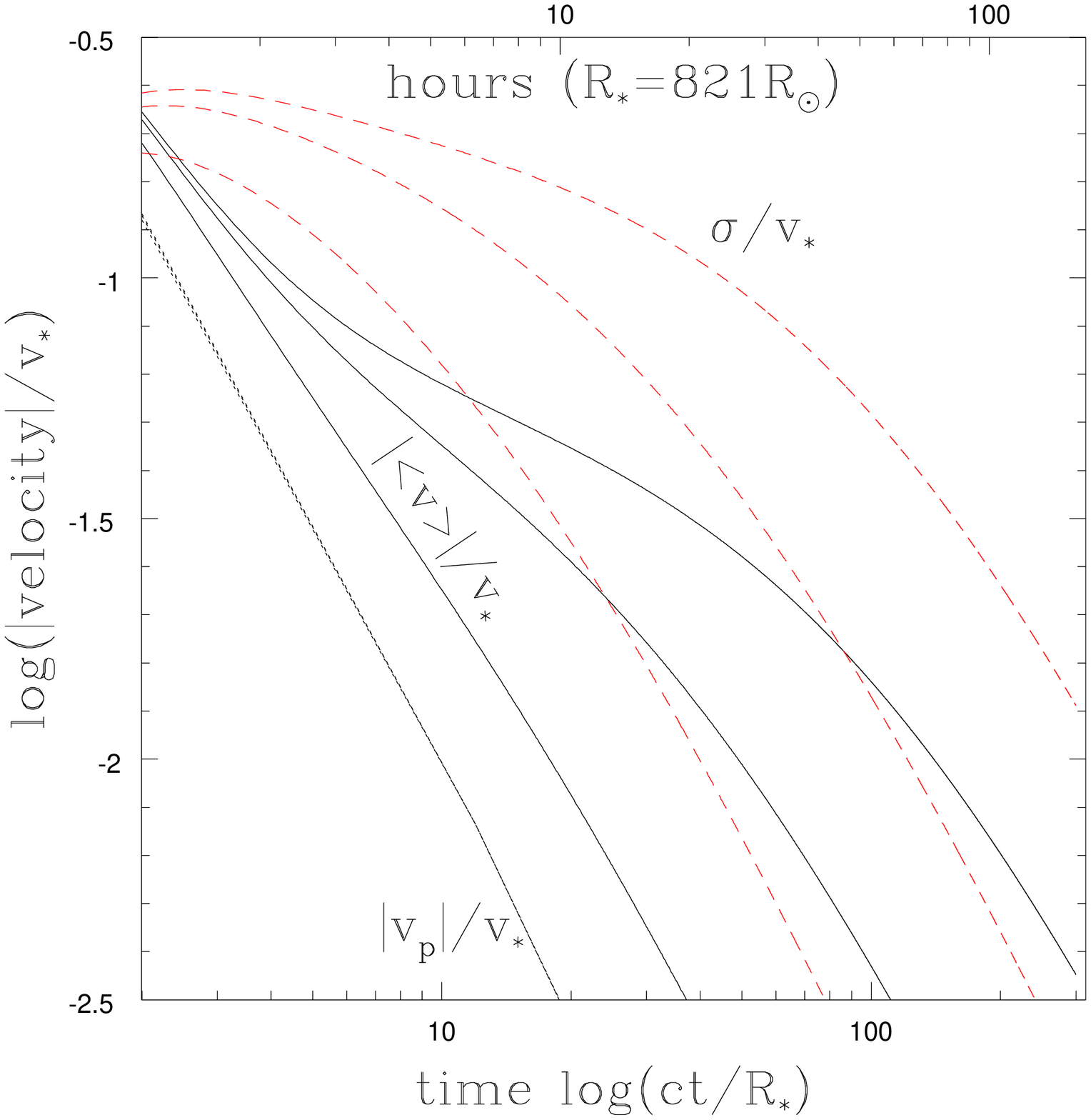}
\caption{
  Evolution of the line peak velocity, $|v_p|/v_*$ (dotted), mean velocity, $\langle v \rangle/v_*$ (solid) and velocity
  dispersion, $\sigma/v_*$ (dashed) for shock velocities of $\beta=v_s/c=0.01$, $0.03$ and $0.1$. Time is in
  units of the stellar light crossing time $R_*/c$ with a conversion to hours for our fiducial stellar radius
  at the top. The line peak and mean velocity are both blue shifted.  Higher shock speeds reduce the mean
  velocity and velocity dispersion but have little effect on the velocity of the peak unless the shock velocity
  is still higher than the velocities shown here.
  }
\label{fig:vbar}
\end{figure}

\section{The Evolution of the Line Fluxes}

In this section we develop a simple model for the time evolution
of the line flux from the flash ionized wind of a star and then apply it to iPTF13dqy.  
We assume an initial shock break out that creates an outward moving shell
of ionizing radiation.  We start by assuming that the photoionizing
flux is high enough to maintain complete ionization and then consider
the evolution as the photoionizing flux decreases.  For simplicity, we consider wind
densities below the point where the shock breaks out of 
the wind rather than the stellar surface (see, e.g.,
\citealt{Chevalier2011}). The expanding
radiation shell instantaneously ionizes the CSM as 
it expands.  We will consider time scales on which the 
finite thickness of the radiation shell can be ignored.
We assume that ionizing photons produced by recombination
are locally reabsorbed (case B) and the nature of the
temporal evolution allows us to ignore effects such as
the production of hydrogen ionizing photons by helium
recombinations.  In fact, for this basic analysis we
will simply assume that all the electrons are due to 
ionized hydrogen.

We are considering a primary with radius $R_{*} = 1000 R_{3}R_\odot$ 
and mass $M_{*} = 10 M_{*10} M_\odot$ producing a $\rho \propto r^{-2}$
wind of velocity $v_{w} = 100 v_2$~km/s and mass loss rate rate 
$\dot{M} = 10^{-4}\dot{M}_4 M_\odot$/year.  For our 
concrete examples, we use the $15M_\odot$ progenitor 
model from \cite{Woosley2002}, which has $R_{*}=821 R_\odot$ and
$M_{*}=12.6 M_\odot$ when it explodes.  We assume
it explodes with energy $E_{SN} = 10^{51} E_{51}$~erg
and has an ejecta mass of $M_e = 10 M_{e10} M_\odot$.  For
our concrete example we adopt $E_{51}=1$ and 
$M_e = 11.2 M_\odot$ (i.e., forming a $1.4M_\odot$
neutron star).

It is useful to normalize recombination rates by the rate
for an infinite wind extending from the stellar surface,
\begin{equation}
    \Gamma_\infty \equiv { \alpha_R \dot{M}^2 \over 4 \pi v_{w}^2 \mu^2 m_p^2 R_{*} }
    \label{eqn:gammainf}
\end{equation}
where $\alpha_R$ is the recombination (or a related) rate and $\mu m_p$
is the relevant mean particle weight.  Exactly what to use for $\mu$
depends on the ionization state of the gas.
Since stellar winds generally have velocities comparable
to the escape velocity of the star (e.g., \citealt{Lamers1999}), we can set the 
wind speed to be $v_{w}^2  = \xi 2 G M_{*}/R_{*}$ where 
$\xi\simeq 1$ is a dimensionless number. For our fiducial
model, we set $\xi = 1$, which makes $v_{w} = 76$~km/s.  
In this formulation, the recombination rate is
\begin{equation}
 \Gamma_\infty =
      { \alpha_R \dot{M}^2 \over 8 \pi \xi \mu^2 m_p^2 G M_{*} }
      =  
 4.3 { \alpha_{13} \dot{M}_4^2 \over \xi \mu^2 M_{*10} } 
    \times 10^{49}~\hbox{s}^{-1}
   \label{eqn:standard}
\end{equation}
where $\alpha_R = 10^{-13} \alpha_{13}$~cm$^3$/s.
Note that there is little freedom to use the wind
speed to adjust the mass loss rate for a wind arising
from a stellar surface -- any change in wind speed is 
balanced by the implied change in stellar radius.
Equating $\Gamma_\infty$ for Case~B H$\alpha$ emission to 
the peak H$\alpha$ flux of iPTF13dqy ($6 \times 10^{50}$ H$\alpha$
photons per second) implies a mass loss rate of  
\begin{equation}
   \dot{M} \simeq 3.7 \xi^{1/2} M_{*10}^{1/2} \alpha_{13}^{-1/2} 
         \times 10^{-4} M_\odot/\hbox{year}.
\end{equation}
We therefore adopt $\dot{M} = 10^{-3.5} M_\odot$/year for
our fiducial models.
\cite{Yaron2017} use a rate $\sim 10$ times higher because they assume a
much larger (stellar) radius, a point we will discuss below. 

The ionized region of the wind is bounded by the expanding
break out pulse and the expanding shock. If the wind is
fully photoionized over the region $R_{s} < R < R_{bo}$, 
the recombination rate is
\begin{equation}
 { \Gamma(t) \over \Gamma_\infty} =  { R_* \over R_s }
             \left( 1 - { R_{s} \over R_{bo} } \right).
      \label{eqn:rate1}
\end{equation}
For the flash ionization problem, the inner and outer radii of
the ionized wind vary with time.  The shock break out pulse
has an ionizing photon rate $Q_{bo} \gg \Gamma_\infty$ and so $R_{bo}$ 
expands by the smaller of the speed of light and
$v_{ion} =  v_w Q_{bo} m_p \dot{M}^{-1}$.  For reasonable
parameters, $Q_{bo}$ is so large that $R_{bo} = R_{*}+c t$.
It does, however, produce a finite number of ionizing photons,
while an infinite $1/r^2$ wind contains an infinite number
of atoms, so the ionization can only extend to some maximum
radius.  This is generally such a large radius that we can
ignore its existence.

The expansion of the shock is more complicated. For 
simple models, we will simply treat the shock velocity
as a constant, $v_s = v_{s0} = 10^4 v_{s4}$~km/s.  
This is reasonable at early times, but at later
times the shock will slow and achieve a self-similar
evolution (\citealt{Chevalier1982}).  For this case, 
we adopt a shock velocity of 
\begin{equation}
    v_s = { v_{s0}  \over 1 + (t/t_{s})^{1/5} },
\end{equation}
which smoothly matches onto the standard self-similar
solution of \cite{Chevalier1982} while removing 
the divergence\footnote{As a practical matter,
we could have simply ignored the fact that the
velocity diverges since the resulting radius 
is perfectly convergent in time.  However, 
this formulation seemed moderately more physical
and was as easy to implement.  Within a few $R_*$
of the primary, there are additional issues in
any case from assuming $v_w$ is constant in the wind
acceleration region.}  
in the velocity ($v_s \propto t^{-1/5}\rightarrow \infty$ 
as $t \rightarrow 0$). The time scale 
\begin{equation}
  t_{s} = { 0.6 E^2 v_{w} \over \dot{M} M_e v_0^5 }
   = 554 {E_{51}^2 v_{w2} \over \dot{M}_4 M_{e10} 
     v_{s0,4}^5}~\hbox{days}
\end{equation} 
is determined by matching the self-similar solution
to the assumed early time shock velocity $v_{s0}$.
Based on the initial expansion rate of the
photospheric radius found by \cite{Yaron2017}, 
$v_{s0} \simeq 20000$~km/s, so if we adopt the
wind parameters from above and $E_{51}=1$ we find
that $t_s=3.8$~days.  The shock radius then evolves as 
\begin{eqnarray}
  R_s(t) &= &R_{*} + { 5 \over 12} t_{s} v_{s0}\times\\
     &&\left[
    3 \hat{t}^{4/5}_{s} - 4 \hat{t}^{3/5}_{s} +
   6\hat{t}^{2/5}_{s}-12\hat{t}^{1/5}_s
    \label{eqn:rshock1}
   + 12 \ln\left(1+\hat{t}_{s}^{1/5}\right)\right] \nonumber
    \label{eqn:modrshock}
\end{eqnarray}
where $\hat{t}_{s}=t/t_{s}$.  As a short hand that
encompasses both solutions we can describe the shock
radius as $R_s(t) = R_{*} + \langle v \rangle t$
where the mean velocity $\langle v \rangle$ is simply a constant or
the more complex expression implied by Equation~\ref{eqn:rshock1}.

Given these definitions, Equation~\ref{eqn:rate1} becomes 
\begin{equation}
 { \Gamma (t)\over \Gamma_\infty}  = 
       \left[ 1 - {\langle v \rangle \over c } \right]
       \left[{ R_{*}  \over R_{*} + \langle v \rangle t }\right]
      \left[ { t \over t + t_*} \right]
\end{equation}
where $t_*=R_*/c$ is the light travel time across the star.
The competition between the addition of newly ionized material
by the expanding radiation front and the absorption of ionized
material into the shocked layer leads to a peak in the
recombination rate.
For the case where $\langle v \rangle = v_{s0}$ is constant,
the recombination rate peaks at 
\begin{equation}
  { \Gamma_{peak}\over \Gamma_\infty}  
     = { 1 - v_{s0}/c \over 1 + v_{s0} /c }
\end{equation} 
when $t_{peak} = R_*/\sqrt{c v_{s0}}$. The peak
recombination rate is close to that for a fully
ionized wind, with $\Gamma_{peak}/\Gamma_\infty=0.97$,
$0.94$ and $0.88$ for $v_{s0}=5000$, $10000$ and
$20000$~km/s, respectively.  For these same 
velocities, the peak occurs at $t_{peak}=7.8$,
$5.5$ and $3.9t_*$.

The observer does not see this light curve
because of light travel times. Light emitted at radius 
$r$ and polar angle $\theta$ is delayed by $r(1-\cos\theta)/c$
relative to emission along the line of sight to the star
($\theta=0$).  The observed radius of a shell expanding at 
velocity $v$ and observed at time $t_o$ is
\begin{equation}
   R(v) = { R_* + v t_o \over 1 + v(1-\cos\theta)/c },
  \label{eqn:bubsize}
\end{equation}
where an observer first receives emission from the 
surface at time $(R_*/c)(1-\cos\theta)$.  This must
be accounted for when $t_o < 2R_*/c$.  To
compute the evolution of the recombination rates, one
can either integrate between $R_{s} = R(v_s)$ and
$R_{bo}=R(c)$, or, the problem 
can be expressed in parabolic coordinates, one of which 
simply corresponds to the time delay (see Appendix~\ref{sec:parabaloid}). 
For a decelerating shock, a reasonable approximation
is to use $v_s = (R_s-R_*)/t=\langle v \rangle$, neglecting any 
velocity changes over the time $R_s/c$ it takes light to
cross the shock.

In particular, for a constant shock velocity, 
the integrals can be done analytically to yield
\begin{equation}
 { \Gamma (t) \over \Gamma_\infty} 
    = \left[ 1 - { v_{s0} \over c } \right] 
      \left[{ R_{*}  \over R_{*} + v_{s0} t } \right]
      \left[ { t^2 \over 4 t_* \left( t+ t_*\right) }\right]
   \label{eqn:delay1}
\end{equation}
for $t < 2t_*$ and 
\begin{equation}
 { \Gamma (t) \over \Gamma_\infty} = 
       \left[ 1 - { v_{s0} \over c } \right]
       \left[  { R_{*}  \over R_{*} + v_{s0} t }\right]
      \left[ { t - t_* \over t +t_*  }\right]
   \label{eqn:delay1b}
\end{equation}
for $t > 2t_*$. The light curve peak now occurs at 
\begin{equation}
   t_{peak} = { R_* \over \sqrt{c v_{s0}} }
 \left[ \sqrt{2} \left(1+ { v_{s0}\over c}\right)^{1/2}
       + \left( { v_{s0} \over c }\right)^{1/2}
  \right]
   \simeq { \sqrt{2} R_* \over \sqrt{ c v_s} }.
   \label{eqn:delay2}
\end{equation}
Adding the time delays smooths and delays the peak
of the emission.  The peak recombination rate,
which has no useful analytic form, drops to
$\Gamma_{peak}/\Gamma_\infty=0.72$, $0.64$ and $0.55$
for $v_{s0}=5000$, $10000$ and $20000$~km/s,
and the time of the peak lengthens to 
$t_{peak} = 12.0$, $8.9$ and $6.7t_*$.

These results ignore any shadowing effects of the 
expanding shock.  If we shadow the region 
with a projected radius of $R_{s}$ behind the
shock, the factor of $1-R_{s}/R_{bo}$ in 
Equation~\ref{eqn:standard} is replaced by
\begin{equation}
 { 1 \over 2} \left[
   1 - {R_{s} \over R_{bo}}
     + { 1 \over 2 } \cos^{-1} {R_{s} \over R_{bo} }
     - { 1 \over 2 } { R_{s} \over R_{bo}} 
       \left( 1 - { R_{s}^2 \over R_{bo}^2 } \right)^{1/2}
   \right].
\end{equation}
This does not include the effects of light travel times
to the observer.
When $R_{bo} \rightarrow R_{s}$, exactly half the 
emission is shadowed and the factor equals $1/2$.
It then increases to $(4+\pi)/8 \simeq 0.89$ as 
$R_{s}/R_{bo} \rightarrow 0$.
In practice, shadowing does not play an important
role in the evolution of the line flux except at
very early times.  For example, for 
$R_*=1000R_\odot$ and $v_s=10^4$~km/s, the factor 
changes by only 3\% ($0.861$ to $0.889$) between $0.1$
and $10$~days. For $R_* = 4000R_\odot$, it changes
by 11\% ($0.798$ to $0.888$), mostly over the first
few hours. Shadowing is important for the shape
of the line profiles, as we discuss in \S3.

This analysis shows that the time of the peak emission
(Equation~\ref{eqn:delay2})
should be a good measure of the radius of the progenitor
star. There is a weak dependence on the shock velocity,
but this can be constrained either by spectra of the
shocked emission or the initial photospheric expansion
rate.  This assumes that the tracer is fully ionized
at the time of the peak, which will generally be true
of hydrogen (see below).  The observed peak recombination rate
is somewhat lower than for the fully ionized wind,
but estimates of the wind density,
$\dot{M} \propto (\Gamma_{\infty}/\Gamma_{peak})^{1/2}$,
will be little affected given other uncertainties.

This brings us to the location of the CSM around iPTF13dqy. 
\cite{Yaron2017} use a size of roughly $1900 R_\odot$,
significantly larger than the expected radius of
a red supergiant.  For an effective temperature
of $T=3500 T_{35}$~K, the stellar radius is 
$R_{*} = 860 L_5^{1/2} T_{35}^{-1/2} R_\odot$ for
a luminosity of $L = 10^5 L_5 L_\odot$, quite
consistent with our fiducial model.
Figure~\ref{fig:lc} shows the evolution of the recombination
rate including light travel times but ignoring shadowing
(i.e., Equations~\ref{eqn:delay1} and \ref{eqn:delay1b}) as a 
function of $R_*$ for a fixed shock velocity of 
$v_{s0}=10^4$~km/s.  We compare to the observed H$\alpha$ 
emission from iPTF13dqy, normalizing the model 
using an unweighted fit to the 
H$\alpha$ measurements. 
The rise time of the H$\alpha$ emission is far
too slow for the star in our fiducial model. In
fact, the rise time is so slow, that it requires
an object with $R_* \simeq 4000 R_\odot$, even
larger than used by \cite{Yaron2017}.  This
result is quite robust against changes in the 
shock velocity, increasing by 33\% if
we double the shock velocity and decreasing by 26\%
if we halve the shock velocity.  
The dense material 
surrounding iPTF13dqy had to have been in a shell
rather than in a wind extending from the surface
of the star.      

\cite{Yaron2017} mention the possibility of moving the
inner edge of the wind off the surface of the star to
make a shell, but do not consider it in any detail.
Figure~\ref{fig:lc2} shows the consequences of changing
the geometry of the wind.  The wind is now a shell
extending from either $R_*$ or $5R_*$ to either
$10R_*$ or $100R_*$. The case of a wind extending 
from $R_*$ to $100R_*$ is fairly similar to an
infinite wind. With our fiducial $R_*=821 R_\odot$, an 
inner edge of $5 R_* \simeq 4100 R_\odot$ 
reproduces the rise of the light curve well, as
we would expect from Figure~\ref{fig:lc}.  
Truncating the wind at $10 R_*$ cannot reproduce
the late time H$\alpha$ flux.  However, 
$10 R_* \simeq 6 \times 10^{14}$~cm is the scale on
which \cite{Yaron2017} propose to truncate the 
wind, and $100 R_* \simeq 6 \times 10^{15}$~cm
is the scale on which \cite{Yaron2017} argue
the wind density must be far lower than closer
to the star.  

It is clear, however, that the dropping production
rate of ionizing photons must also
play a role since 
the higher ionization potential lines seen in
iPTF13dqy evolve more rapidly than the lower
ones.  As the emission rate $Q(t)$ of ionizing
photons drops, there are eventually too few
to balance recombination in the wind.  Since
the wind is densest close to the shock, we
can model this by assuming a fully ionized
region extending from the shock to the radius
where the integrated recombination rate 
equals $Q(t)$.  As this radius retreats, 
previously ionized material simply begins
to recombine.  Additionally, material newly
photoionized by the still expanding shell
of radiation from the shock break out begins
to recombine immediately afterwards. Thus,
we need to find the radius $R_{pi}(t)$ 
where recombination balances ionization,
\begin{equation}
  Q(t,E_i) = X \Gamma_\infty { R_* \over R_{s}(t) }
  \left( 1- { R_{s}(t) \over R_{pi}(t)} \right).
  \label{eqn:pi}
\end{equation}
This depends on the abundance (by number) of the 
atom $X$ and the ionization energy $E_i$
of interest.  A more detailed calculation 
should also consider the material between the
photosphere and the shock front.

We use $X=1$, $0.1$ and $5.4\times 10^{-4}$ 
for hydrogen, helium and oxygen, respectively. Using
$X=1$ for hydrogen, and simply scaling Equation~\ref{eqn:pi}
with $X$ corresponds to the assumption that all  the
available electrons are associated with the ionization
of hydrogen and that hydrogen remains fully ionized 
until helium and oxygen are effectively neutral. These
simplifications are quite reasonable.   
We considered ionization
energies of $13.6$~eV (hydrogen), $24.6$~eV (HeI ionization)
$54.4$~eV (HeII and OIII ionization), $77.4$~eV (OIV ionization)
and $113.9$~eV (OV ionization).  Note that because the HeII
and OIII ionization energies are so similar, we should use
$X=0.1$ for helium in both cases.  

We estimate $Q(t)$ using a simple model for the evolution
of the estimated black body temperature, photospheric
radius and luminosity for iPTF13dqy from \cite{Yaron2017}.
Using a black body spectrum likely overestimates the number
of ionizing photons for a given effective temperature but
seems adequate for this exploration.
We find that the temperature of iPTF13dqy is well-approximated by 
\begin{equation}
  T_{BB}(t) 
     \simeq 2500 + 6940 \left( { 10~\hbox{days} \over t } \right)^{1/2}~\hbox{K}
\end{equation}
and the photospheric radius is well-approximated by
\begin{equation}
  R_{BB}(t)
     \simeq 10^{13.62} + 10^{15.24} { t \over t + 10~\hbox{days}}~\hbox{cm}.
\end{equation}
The largest discrepancy is for the first epoch at 4~hours
which could be rectified by slightly adjusting the time
zero point. However, the exact properties at this epoch
are not crucial to our discussion. Note that the derivative
of $R_{BB}(t)$ is not expected to be a good estimate of the
shock velocity, so there is no need to reconcile this 
empirical fit with Equation~\ref{eqn:modrshock}. The luminosity is 
simply $L = 4 \pi \sigma R_{BB}^2 T_{BB}^4$.  This 
allows us to compute the evolving number of photoionizing 
photons above energy $E_i$,
\begin{equation}
   Q(t,E_i) = { 15 L(t) \over \pi^4 k T(t) } \gamma\left({ E_i \over k T(t)}\right)
     \label{eqn:have}
\end{equation}
where
\begin{equation}
  \gamma(u) = \int_u^\infty du u^2(\hbox{e}^u-1)^{-1}.
\end{equation}
We can then combine this with Equation~\ref{eqn:pi} to 
determine $R_{pi}(t)$, and then invert it to determine 
the time $t_{pi}(R)$ when material at radius $R$ 
ceases to be fully photoionized. 

Figure~\ref{fig:spacet} shows the resulting evolution of 
$R_{pi}$ for these five cases.  Because the number of 
ionizing photons is dropping exponentially as the 
photosphere cools, $R_{pi}$ evolves very quickly.  
Over a very short period it drops from a very large
radius to be very close to the shock radius and 
then asymptotically tracks the expanding shock.
For the particular parameters used in Figure~\ref{fig:spacet},
it is no longer possible to 
ionize HeII, OIII, OIV or OVI after roughly 12 hours.
After three days
it is no longer possible to ionize HeI, and after
roughly 30 days it is no longer possible to 
ionize HI.  This is in qualitative
agreement with the spectral evolution found by
\cite{Yaron2017}, where the oxygen emission lines
vanish in less than a day while the H$\alpha$
emission persists.  

Given $t_{pi}(R)$, we can then estimate the declining
recombination rate given the recombination time of
\begin{equation}
   t_r= { 4 \pi v_w m_p R^2 \over \alpha \dot{M} }
    = 0.019 { v_{w2} R_3^2 \over \alpha_{13} \dot{M}_4}~\hbox{days}.
\end{equation}
This is also shown in Figure~\ref{fig:spacet} for 
our standard parameters.  If
the recombination time is short compared to the present
time, the recombination rate will drop rapidly, and
{\it vice versa}.  If $\Delta t = t - t_{pi}(R)$  
is the time since photoionization ceased, the
recombination rate of helium and oxygen declines
as $\exp(-\Delta t/t_r)$ while that of hydrogen
declines as $(1+\Delta t/t_r)^2$.  The difference
arises from the assumption that the electrons
are all associated with hydrogen, so that the
recombination time at any radius is constant for helium and
oxygen but steadily becomes longer as the
hydrogen ionization fraction drops.
For the final calculations, we combine this model
for the effects of the declining numbers of 
ionizing photons and Equation~\ref{eqn:rshock1}
for the expansion of the shock with
Equation~\ref{eqn:parab1} to include the 
effects of the propagation delays.  

Figure~\ref{fig:recomb1} shows the fiducial model for 
the recombination rates to OV, OVI, OIII/HeII, HeI 
and H.  The outer envelope is the same as for Figure~\ref{fig:lc} 
but for a shock that is slowing down as it expands. 
When the production rate of the necessary
ionizing photons drops below that needed
to ionize the wind outside the shock front,
the material which has the shortest recombination
times recombines first, leading to a rapid
initial drop in the recombination rates followed
by a tail produced by
material with longer recombination times and
the continuing photoionization of new material
by the still expanding radiation front.  Like
the H$\alpha$ comparison in Figure~\ref{fig:lc},
everything occurs earlier in this fiducial 
model than was observed for iPTF13dqy -- the
evolution of all the lines requires that the 
material dominating the emission is detached
from the stellar surface. 

\begin{figure*}
\centering
\includegraphics[width=0.9\textwidth]{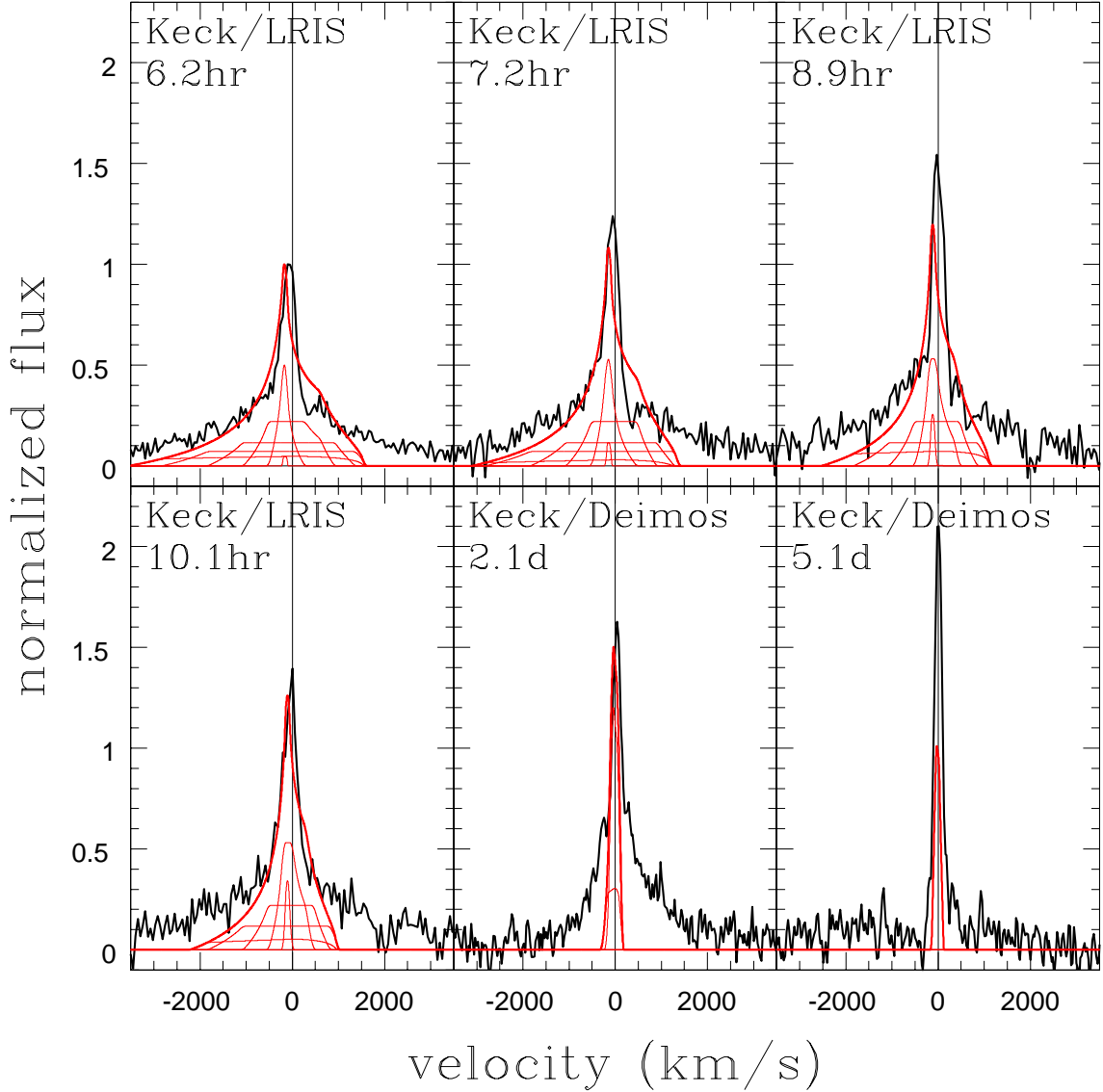}
\caption{ The H$\alpha$ emission line (black curves) of iPTF13dqy in velocity space with the flux
  normalized to a peak of unity in the first epoch (upper left) and then following its evolution
  to $8.8$~days.  The heavy red curve shows a radiative acceleration plus recombination model for
  the evolution of the line for a $\rho \propto 1/r^2$ wind extending from $R_*$ to $10^4 R_*$.
  The light solid lines show the contributions from radial regions of $1.67$-$2$, $2$-$2.5$,
  $2.5$-$3.33$, $3.33$-$5$, $5$-$10$ and $10$-$10^4R_*$ around the star, defined so that the
  recombination line fluxes of each region when fully ionized equal $10\%$ of the flux of a
  fully ionized wind extending from the star's surface.  Regions closer to the star have
  broader velocity profiles.  Regions make no contribution when either fully inside the
  observed shock radius or outside the expanding photoionized region.  Time delays and
  shadowing of the CSM by the expanding shock are included and lead to the line asymmetries.
  The velocity is the width of the top hat used to model instrumental resolutions.
  }
\label{fig:spectra}
\end{figure*}

\section{Radiative Acceleration: Broad Line Wings Without Thomson Scattering}

\label{sec:acceleration}

\cite{Yaron2017} argue that an eruption is needed to produce a CSM dense
enough to explain the broad wings of the narrow emission lines seen for the first
several days using Thomson scattering.  Here we note that radiative 
acceleration of the wind by the shock break out pulse from an RSG will also naturally
produce such line profiles independent of the wind density.  The 
break out radiation has energy $E_{bo}$ and hence momentum $E_{bo}/c$.
For a Thomson optically thin wind, an electron at radius $R$ will capture 
$ E_{bo}\sigma_T/ 4 \pi c R^2$ of this momentum. For a Solar ratio of
hydrogen and helium, the mean mass per electron is $\alpha m_p$ with
$\alpha = 1.17$, leading to a velocity of 
$E_{bo}\sigma_T/ 4 \pi c \alpha m_p R^2$.  We can safely assume that
the medium is ionized because photoionization cross sections are
much larger than Thomson and recombination rates are much slower
than Thomson scattering rates.  After the
break out radiation pulse has passed, a wind starting with constant
velocity $v_w$ now has the velocity profile
\begin{equation}
   v(r) = v_w + v_* \left( R_* \over r \right)^2
\end{equation}
scaled to the velocity $v_*$ at the stellar surface.  This will trigger
a hydrodynamic response, but we can view the velocity profile as 
fixed for the short period of time we consider here. That radiative
acceleration produces such velocity profiles has been considered
previously, but always in the context of line profiles 
modified by Thomson scattering in Type~IIn SNe at much later
times (e.g., \citealt{Chugai2001}, or
\citealt{Huang2018} more recently).  There seems 
not to have been a discussion of this radiative acceleration effect on the 
recombination line profiles independent of the Thomson optical
depth (it may implicitly be included in \cite{Chugai2001}, but the focus is on
the Thomson scattering).   It certainly has not been considered
in these early phases where light travel times are also relevant.  

Using the $n=3/2$ results from \cite{Matzner1999}, the break out energy is roughly 
\begin{equation}
E_{bo} \simeq 6 \kappa_e^{-0.87} \hat{\rho}^{-0.086} E_{51}^{0.56}
        M_{e10}^{-0.44} R_3^{1.74} \times 10^{48} \hbox{erg}
\end{equation}
where $\kappa_e$ is the opacity in units of $0.34$~cm$^2$/g
(i.e., Thomson), $\hat{\rho}^{-0.086} \simeq 1$, and the remaining quantities 
were defined in \S2.   This sets a velocity scale of  
\begin{eqnarray}
   v_* &= &{ E_{bo} \sigma_T \over 4 \pi c \alpha m_p R_*^2}  \\
      &\simeq &11000 \kappa_e^{-0.87} \hat{\rho}^{-0.086} E_{51}^{0.56}
        M_{e,10}^{-0.41} R_{*3}^{-0.26} \hbox{km/s}, \nonumber
       \label{eqn:vrad}
\end{eqnarray}
which is remarkably high. The acceleration period lasts roughly
an hour for a red supergiant, and we ignore this initial transient
phase.  Since the SN shock is essentially expanding at close
to $v_*$, the temporal window for seeing velocities close to $v_*$
is short.

Given the
velocity profile established by the radiative acceleration,
and the observed (i.e., including time delays) locations of the
shock and the radiation pulse (Equation~\ref{eqn:bubsize}), 
we can compute the expected
shape of the emission line profile as a function of time.
Here we include the shadowing of the CSM by the expanding
shock, as this has significant effects on the line shapes.
Figure~\ref{fig:vbar} shows the evolution of the line
peak, $v_p$, the mean velocity, $\langle v \rangle$, and
the dispersion in velocity, $\sigma$, for a wind speed
$v_w=0$.  The velocities are scaled by $v_*$, and both the
line peak and the mean velocity are blue shifted.  If we
measure time in units of the stellar light crossing time,
$ct /R_*$, the profiles depend only the shock speed 
relative to the speed of light, $\beta=v_s/c$, and
we show results for $\beta=0.01$, $0.03$, and $0.1$.
The mean and the dispersion in
velocity decrease as the shock speed increases because
the fastest moving parts of the CSM are shocked earlier.  
The Doppler shift of the line peak is nearly 
independent of $\beta$.
Figure~\ref{fig:vbar} also shows the conversion to
hours assuming our fiducial stellar radius.  For a 
decelerating shock, Figure~\ref{fig:vbar} can be
interpreted using the mean shock velocity, 
$\beta = (R_s(t)-R_*)/c t=\langle v\rangle/c$. This (safely) ignores
deceleration of the shock on a light crossing time.

Figure~\ref{fig:spectra} shows the expected line evolution for a
constant velocity wind extending from $R_*$ to $10^4 R_*$
with $R_*=821R_\odot$, $v_w=76$~km/s, $v_s=20000$~km/s and
$v_* \simeq 11000$~km/s based on Eqn.~\ref{eqn:vrad}.  The times are 
chosen to match the Keck spectroscopic observations of iPTF13dqy by
\cite{Yaron2017}. They are smoothed with a $100$~km/s top hat
to approximate the effects of finite spectral resolution,
and normalized to a peak of unity in the first epoch.
To illustrate the regions responsible
for the line shape, we show the contributions from radial
regions that when fully ionized will produce 10\% of the 
recombination radiation of a fully ionized wind. The chosen
annuli are $1.67$-$2$, $2$-$2.5$, $2.5$-$3.33$, $3.33$-$5$, 
$5$-$10$ and $10$-$10^4R_*$.  Regions closer to the star 
are swept over by the shock very quickly and have such 
broad velocity widths that they would be very difficult to
separate from the continuum emission. 

At the earliest epoch, the line flux is dominated by very
broad wings from regions close to the star, with little 
contribution from the outermost annulus.  The narrow peak
is blue-shifted because the most redshifted material is
shadowed by the shock and because of the near/far
asymmetry in the observed photoionized region (Eqn.~\ref{eqn:bubsize}).
This also leads to a
line asymmetry with more blue than red flux at higher velocities.
The lines narrow and the blue shifts diminish as the shock passes 
over the most accelerated material and the photoionization front 
reaches larger radii.  The
peak recombination flux, as discussed in \S2, occurs at 
intermediate times.  Note that these line profiles
are independent of the Thomson optical depth of the CSM, 
they are purely due to the radiative acceleration.  

For comparison, we show the spectra of the H$\alpha$ region
from \cite{Yaron2017}. We obtained the spectra of iPTF13dqy
from the WISeREP (\citealt{Yaron2012}) repository.  We
corrected for redshift, using the line center in the 
Keck/Deimos $5.1$~day spectrum to define zero velocity.
The mean fluxes in windows adjacent to the line
were used to subtract a linear continuum.  
These windows are sufficiently far removed from the line 
center that they should be little contaminated even by very 
broad line wings.  This model for the continuum works
reasonably well, although there are several epochs where
the continuum is not very flat.

The models qualitatively reproduce the evolution of the spectra 
even though we made no adjustments to the parameters.  The
observed line profiles are somewhat less asymmetric and the
early-time peaks are less blue-shifted ($\simeq 80$~km/s versus
$\simeq 150$~km/s at 6.2~hr), and either a larger $v_*$ or
a lower $v_s$ would help to keep the lines broader for longer.
Allowing the shock to decelerate would also keep the lines
broader for a longer period of time.  One could also increase
$v_*$ while having less material near the star, a solution
suggested by the slow rise time of the H$\alpha$ emission.
Since in \S\ref{sec:wind} we will propose a very different
explanation and more geometrically complex solution, we
did not attempt to optimize the model.

Many of these effects would have a similar effect on the
line profiles produced by Thomson scattering.
For example, the optical depth is
largest, and so the scattering wings broadest, at early times.
The near/far asymmetry produced by the time delays 
and the shadowing by the shock would again produce blue/red
asymmetries in the line profiles.  The Thomson optical depth of
the CSM will scale as $\tau_0 R_*/R_s$ where $R_s$ is the shock
radius, and the scattering wings will have a width 
$\sim v_e \tau$ assuming a random walk in 
velocity space (i.e., $N \sim \tau^2$ scatterings) 
with $v_e \simeq 10^3$~km/s for electrons with $T \simeq 20000$~K.  
The optical depth changes little over the first four
epochs, dropping by only 20\% between 6.2 and 10.1~hours
for our fiducial parameters, but then falling rapidly
to $0.24\tau_0$ and $0.11\tau_0$ by 2.1 and 5.2~days. As
the optical depth drops, the lines become narrower.

\begin{figure}
\centering
\includegraphics[width=0.45\textwidth]{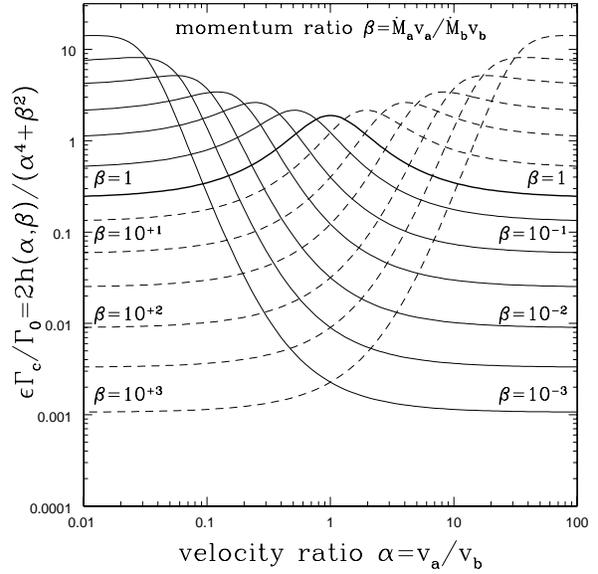}
\caption{ The efficiency factor $\Gamma_c/\Gamma_b$ from Equation~\ref{eqn:factor}
   as a function of the relative wind speeds
  $\alpha=v_a/v_b$ for a range of wind momentum ratios $\beta= \dot{M}_a v_a/\dot{M}_b v_b$
  for recombination in a thin shock created by
  colliding winds relative to recombination in the winds of the binaries starting at a radius
  equal to the binary separation.
  }
\label{fig:factor}
\end{figure}

\begin{figure}
\centering
\includegraphics[width=0.45\textwidth]{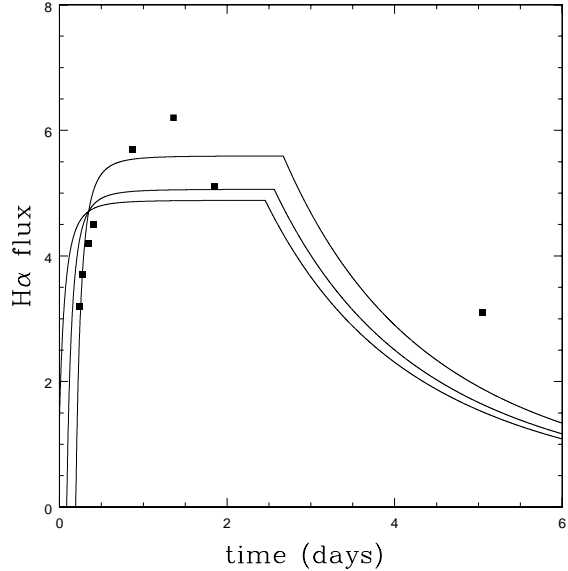}
\caption{ The recombination light curves for a $\beta=1$ colliding wind with $D/2=4000R_\odot$,
  $R_*=821R_\odot$ and $v_{s0}=10^4$~km/s.  The three curves correspond to putting the secondary
  along the line of sight to the primary (fastest rise), perpendicular to the line of sight (middle)
  and on the far side of the primary (slowest rise).   No other light travel time effects are
  included.  The points are the observed H$\alpha$ light curve from \protect\cite{Yaron2017}, and
  the curves are normalized to fit the equally weighted observations.
  }
\label{fig:newcurve}
\end{figure}

\section{A Colliding Wind}
\label{sec:wind}

If we take the LBT and X-ray evidence at face value, RSGs do not have pre-SN outbursts,
so we require an alternate explanation.  While radiative acceleration can naturally
explain the broad line wings without the need for a medium optically thick
to Thomson scattering, the recombination flux (Equation~\ref{eqn:standard}) 
still requires a high density CSM.  To
avoid the need for a pre-SN eruption, we need a mechanism to increase the 
recombination rate given a mass loss rate normally associated with an RSG.
This requires a means of
producing a denser medium than expected for such a wind, with a shell-like
structure to explain the rise time and the lack of radio/X-ray emission
at late times. 

We expect a large fraction of SNe to occur in binaries, where the binary
companion will typically be a hotter, main sequence star (see, e.g.,
\citealt{Kochanek2009}).  Such stars have relatively low $\dot{M}$, high
velocity winds, so photoionizing the wind of the secondary also 
cannot produce the observed line fluxes.
However, the secondary also modifies the winds.  In particular, there
is a shocked boundary layer created by the collision of the winds
from the two stars.   In many Wolf-Rayet binaries, the shocked material
cools to the point of dust formation (e.g., \citealt{Tuthill1999}
\citealt{Monnier1999}, \citealt{Monnier2002},
\citealt{Daugherty2005}, \citealt{Tuthill2008}, \citealt{Williams2009}), 
which means that the density
must far exceed the density a wind would have at such a distance from 
an isolated star.  Since recombination is a $n^2$ process, this can
greatly enhance the recombination rate over expectations for a 
smoothly expanding wind in the same way that clumping in stellar winds
can significantly enhances line strengths for a given mass
loss rate (e.g., \citealt{Puls2008}).  While we focus on colliding winds because there is
an elegant analytic model, hot secondaries can produce similar effects
without any wind because of the HII region that they form in the 
wind of the primary (see the $\alpha$ Sco models of 
\citealt{Braun2012} and the short discussion in Appendix~\ref{sec:h2region}).

\cite{Canto1996} present an analytic solution for the geometry and surface density of an
axisymmetric colliding wind based on mass and momentum conservation and assuming that
efficient cooling makes the shocked region ``infinitely'' thin.  Star a (the primary)
is at the origin and star b (the secondary) lies at a distance $D$ along the $z$ axis.  Relative to star a
geometry provides the relation
\begin{equation} 
       R(\theta_a) = D \sin\theta_b \csc\left(\theta_a+\theta_b\right) = D f(\theta_a)
\end{equation}
where $\theta_a$ is a spherical polar angle centered on star a with $\theta_a=0$
pointing to star b and $\theta_b$ is a spherical polar angle centered on star 
b with $\theta_b=0$ pointing to star a.  Mass and momentum conservation lead
to the constraint that
\begin{equation}
      \theta_b \cot\theta_b = 1 + \beta \left( \theta_a \cot\theta_a -1 \right)
\end{equation}
where $\beta = \dot{M}_a v_a / \dot{M}_b v_b$ is the ratio of the wind momenta.
The surface density in the shock is then $\sigma(\theta_a) = \sigma_0 g(\theta_a)$
where $\sigma_0 = \dot{M}_a/2\pi \beta v_a D$, and $g(\theta_a)=A/B$ with
\begin{eqnarray}
  A &= & \sin\left(\theta_a+\theta_b\right)\csc\theta_a \csc\theta_b  \times \nonumber \\
      &  & \left[ \beta\left(1-\cos\theta_a\right) + \alpha\left(1 - \cos\theta_b\right)\right]^2,  \\
  B^2 &= & \left[\beta\left(\theta_a-\sin\theta_a\cos\theta_a\right) + \left(\theta_b-\sin\theta_b\cos\theta_b\right) \right]^2+\
                \nonumber \\
      &  &  \left[\beta\sin^2\theta_a-\sin^2\theta_b\right]^2
   \nonumber
\end{eqnarray}
and $\alpha=v_a/v_b$.

We can estimate the recombination rate associated with the wind interface
as follows.  First, we assume that the layer has constant fractional
thickness, $\Delta = \epsilon R(\theta_a) = \epsilon D f(\theta_a)$,
relative to the distance from the primary.  This makes the gas density
$\rho = \sigma_0 g(\theta_a)/\Delta$.  We also need the area element
of the shock surface,
\begin{equation}
   d A  = 2 \pi D^2 f(\theta_a)\sin(\theta_a)
          \left[ f(\theta_a)^2 + \left( { df(\theta_a)\over d\theta_a }\right)^2 \right]^{1/2}
            d\theta_a.
\end{equation}
The recombination rate per unit area is $\alpha_R \rho^2/\mu^2 m_p^2 \Delta$, and
the total is
\begin{equation}
  \Gamma_c = { 2 \pi \alpha_R \sigma_0^2 D \over \epsilon \mu^2 m_p^2 }
         h(\alpha,\beta)
    \label{eqn:colrate}
\end{equation}
where
\begin{equation}
   h(\alpha,\beta) = \int d\theta_a { \epsilon D g^2(\theta_a) f(\theta_a) \sin(\theta_a) \over \Delta}
           \left[ f^2 + \left( { df\over d\theta_a }\right)^2 \right]^{1/2}
\end{equation}
which for our assumed scaling of the thickness is
\begin{equation}
   h(\alpha,\beta) = \int d\theta_a g^2(\theta_a) \sin(\theta_a) 
           \left[ f^2 + \left( { df\over d\theta_a }\right)^2 \right]^{a/2}.
\end{equation}
For winds of equal momentum ($\beta \equiv 1$), this becomes
\begin{equation}
   h(\alpha,\beta\equiv 1) = 16 (1+\alpha)^4 \int d\theta_a 
        { \sin^6(\theta_a/2)\tan(\theta_a/2)\over (2\theta_a-\sin\theta_a)^2 }
\end{equation}
which is $h(\alpha,\beta\equiv 1) \simeq 0.12 (1+\alpha)^4$ 
when integrated over the full shock ($0 \leq \theta_1 \leq \pi/2$).

The overall scale can be made clearer by normalizing the recombination rate
to the rate which would be produced by the two stellar winds starting from
a radius equal to the binary separation $D$, 
\begin{equation}
  \Gamma_D ={ \alpha_R \over 4\pi \mu^2  m_p^2 D }
     \left({ \dot{M}_a^2\over v_a^2} +{\dot{M}_b^2\over v_b^2}\right).
\end{equation}
This is less than the rate for an individual star of radius $R_*$ (Equation~\ref{eqn:gammainf}) 
by $D/R_*$.  Given this scaling, the recombination rate from the wind
collision region becomes
\begin{equation}
  \Gamma_c = \epsilon^{-1} \Gamma_D { 2 h(\alpha,\beta) \over \alpha^4 + \beta^2 }.
  \label{eqn:factor}
\end{equation}
This function is shown in Figure~\ref{fig:factor}.
As required, the result is unchanged if the labels of the two stars are 
reversed ($\alpha \rightarrow \alpha^{-1}$ and 
$\beta \rightarrow \beta^{-1}$).   For equal momentum winds ($\beta\equiv 1$)
this becomes 
\begin{equation}
  \Gamma_c = { 8 \alpha_R (1+\alpha)^4 \dot{M}_a^2 \over \epsilon \pi D \mu^2 m_p^2 v_a^2}
     \int d\theta_a 
        { \sin^6(\theta_a/2)\tan(\theta_a/2)\over (2\theta_a-\sin\theta_a)^2 }
\end{equation}
which when integrated over the shock is 
\begin{equation}
   \Gamma_c \simeq { 0.29 (1+\alpha)^4 \over \epsilon } 
         \left[ { \alpha_R \dot{M}_a^2 \over 4 \pi v_w^2 \mu^2 m_p^2 D } \right]
\end{equation}
where the term in brackets is the same as in Equation~\ref{eqn:gammainf} for a
star of radius $D$. 

If we now equate this to the peak H$\alpha$ flux (see \S2), then
\begin{equation}
  \dot{M}_a \simeq 3.5 v_{a2} \epsilon^{1/2} D_4^{1/2} \alpha_{H\alpha 13}^{-1/2}
         \times 10^{-3} M_\odot/\hbox{year}
\end{equation}
where $D = 10^4 D_4 R_\odot$ places the shock at the right distance to
explain the rise time of the line fluxes, the H$\alpha$ rate is scaled
to $\alpha_{H\alpha} = 10^{-13} \alpha_{H\alpha 13}$~cm$^3$/s 
corresponding to a temperature of roughly $10^4$~K (e.g., \citealt{Draine2011}), and we
have assumed that the primary is a red giant with a slow wind ($v_a = 100 v_{a2}$~km/s) and
the secondary is a hot main sequence star with a lower mass loss rate
($\dot{M} \sim 10^{-10}M_\odot$/year) but a very high wind velocity
($v_b \simeq 10^3$~km/s).  This means that the RSG wind
wind has far more momentum and we have $\beta \simeq 100$ and
$\alpha \simeq 10^{-2}$. Examining Figure~\ref{fig:factor}, we
see that this reduces the dimensionless factor in Equation~\ref{eqn:colrate}
by a factor of 20 and so requires an increase in the mass loss
rate of a factor of 4-5 to get the same recombination rate.

Finally, we assume that the shock collision region cools as it does in
the colliding wind Wolf-Rayet binaries.  The surface density at the
stagnation point (the position of the interface on the axis separating
the stars) is $\sigma_s = 3(1+\alpha)^2 \beta^{1/2} \sigma_0/8$,
so the density is 
\begin{equation} 
    \rho_s = { 3 (1+\alpha)^2 (1+\beta^{1/2})\dot{M}_a \over 16 \pi \epsilon \beta v_a D^2 }.
\end{equation}
Equating the pressure in the layer $\rho_s kT/\mu m_p$ to the incoming
ram pressure from the RSG we find that
\begin{equation}
   \epsilon = { 3 (1+\alpha)^2 \over 2 (1+\beta^{1/2}) } { k T \over \mu m_p v_a^2 } \nonumber \\
            \simeq  0.06 \mu^{-1} v_{10}^{-2} T_3  
\end{equation}
for $\beta=1$ and $\alpha \simeq 0$ with the pre-factor dropping to
$0.01$ for $\beta=10^2$ and $\alpha \simeq 0$ so that  
\begin{equation}
  \dot{M}_a \sim D_4^{1/2} T_3^{1/2} \alpha_{H\alpha 13}^{1/2}
         \times 10^{-4} M_\odot/\hbox{year}.
\end{equation}
While still high, this is now in the regime of RSG winds and
largely avoids the late-time emission limits from \cite{Yaron2017}.  
The required $\dot{M}$ can be driven downwards by including the
photoionization of the winds of the primary by the secondary
(Appendix~\ref{sec:h2region}), which should be similar
to making $\beta$ closer to unity, or by scaling the thickness
$\Delta$ of the interface with radius more slowly than as a
constant fraction of the radius. 

\section{Discussion}

In this paper we introduce three general points about flash
spectroscopy and the potential effects of binarity on 
interpretations of flash spectroscopy.  The first general
point (\S2) is that the time-dependent flux of the low ionization
lines determines the size of the star. In particular, the
flux peaks at $t_{peak} \simeq R_* \sqrt{2/c v_{s0}}$ 
(Eqn.~\ref{eqn:delay2}) where $R_*$ is the stellar radius
and $v_{s0}$ is the shock speed.  The time $t_{peak}$ is a 
trade off between the outgoing radiation pulse from the shock 
break out ionizing more material and the outgoing shock
front running over the ionized material.  The decreasing
temperature of the radiation field eventually cuts off the
line emission, with the emission from high ionization energy 
lines being cut off earlier than low ionization energy lines.  
This means that the hydrogen Balmer lines will be the best
probes of the size of the emission region.

The second general point (\S3) is that flash spectroscopy
observations should always find broad line wings at early times.
For Type~II SNe with
giant progenitors, radiative acceleration produces broad wings 
independent of the Thomson optical depth of the CSM.  
For  
stripped envelope SNe, radiative acceleration is much less
effective because far less energy is associated with the 
shock break out (e.g., \citealt{Matzner1999}), but the 
line-accelerated winds expected for the progenitors of
stripped envelope SNe will intrinsically possess broad 
wings.  Detection of broad wings in flash spectroscopy
of Type~II SNe are thus a test of the predicted energetics
of shock break out radiation.  Since the velocity profile
created by radiative acceleration is $\propto r^{-2}$, the
lines will narrow quite rapidly as the shock front runs
over the fastest material.  

The third general point (\S4) is that many SN occur in 
binary systems (e.g., \citealt{Kochanek2009}) and binaries 
sculpt the winds of the primary through both their winds 
and their production of ionizing radiation. In particular,   
the winds from the two stars collide, producing a boundary
layer separating the two winds (e.g.,
\citealt{Canto1996}).  The shock
collision regions are directly observed in some Wolf-Rayet
binaries (e.g., \citealt{Monnier1999}, \citealt{Monnier2002},
\citealt{Daugherty2005}, \citealt{Tuthill2008}, \citealt{Williams2009}).
More importantly, these systems form dust,
which means that the gas
must cool to produce a very dense, cold layer, shielded from
the harsh radiation environment produced by the Wolf-Rayet
stars.  
Cooling in the shock provides a source of high density material 
that can produce flash spectroscopy recombination rates that 
neither star could produce in isolation.  
It is likely that the sculpting of the CSM by the winds and
photoionizing fluxes of secondary companions explains many
of the deviations seen in X-ray (e.g., \citealt{Dwarkadas2012})
or radio (e.g., \citealt{Margutti2017} or \citealt{Chandra2018}) light curves from the expectations for
a simple $\rho \propto r^{-2}$ wind.  Detailed simulations
are needed to follow the fully-developed
``pinwheel'' shock geometry created by orbital motions
(e.g., \citealt{Stevens1992}, \citealt{Parkin2008}, \citealt{Lamberts2011},
\citealt{Lamberts2012}).

We also discuss these effects for the specific example of iPTF13dqy,
an otherwise normal Type~IIP SNe where \cite{Yaron2017} use their
flash spectroscopy results to argue for a short lived, high
mass loss rate ($\dot{M} \sim 10^{-3} M_\odot$/year) pre-SN
eruption that produced a dense CSM surrounding the star.  We
were driven to search for an alternate explanation because 
at least two lines of evidence indicate that normal Type~IIP
SNe do not have such eruptions:  (1) the     
absence of such events for 4 normal Type~II SNe (\citealt{Kochanek2017},
\citealt{Johnson2017})
in the LBT search for failed supernovae (\citealt{Gerke2015},
\citealt{Adams2016}) and (2) the absence of X-ray emission from
normal Type~IIP SNe (e.g., \citealt{Dwarkadas2014}).    

First, the time evolution of iPTF13dqy's H$\alpha$ 
light curve requires that the dense CSM starts at $\sim 4000R_\odot$ and
so must be detached from the star.
Moreover, the continued H$\alpha$ emission after $\sim 5$~days
requires that the dense CSM extends to the radial scales where 
\cite{Yaron2017} propose that it is truncated.  
Second, the broad line wings are created by radiative acceleration 
and confirm the energy scale of $E_{bo} \sim 10^{40}$~erg
predicted by theoretical models of shock break outs from RSGs.
The broad line wings are not evidence of a CSM optically 
thick to Thomson scattering.  Both the evolution of the
line fluxes and the line profiles appear to require a 
more complicated geometry than a simple wind extending
from the stellar surface and the line fluxes still require
the existence of a denser CSM than a normal RSG wind.
The geometry and the line fluxes are both consistent 
with the progenitor having a relatively normal 
RSG wind which is swept up into a dense, cooling layer by the 
wind of a secondary to produce a much higher recombination 
rate than would be expected for an isolated wind.  Because
the wind densities are much lower than invoked by \cite{Yaron2017}
and the geometry of the dense shell (``paraboloidal'') is different from 
the geometry of the expanding shock (``spherical''), there
is no difficulty staying under the later time limits on
X-ray or radio emission from the system.  

These points also explain the peculiarities of the 
Type~IIb SN iPTF13ast (SN~2013cu, \citealt{Galyam2014}).  
Like iPTF13dqy, flash spectroscopy of iPTF13ast showed broad wings which
\cite{Galyam2014} interpret as being intrinsic to
a pre-SN, fast ($\sim 10^3$~km/s) Wolf-Rayet-like wind
because their model of the wind is Thomson optically
thin and so cannot produce the wings by scattering.
\cite{Galyam2014} cite SN~2008x as an example of a
Type~IIb with a Wolf-Rayet progenitor (\citealt{Crockett2008}), 
but this hypothesis is incompatible with more recent
observations (\citealt{Folatelli2015}).  The best-studied 
progenitors of Type~IIb SN are all yellow supergiants 
(e.g., SN~1993J and SN~2011dh, \citealt{Aldering1994}, \citealt{Maund2011}, 
\citealt{VanDyk2011}) which should have far slower wind
speeds than Wolf-Rayet stars since stellar wind speeds are 
generally comparable to stellar escape velocities. Moreover,
yellow supergiants are unlikely to have 
line driven winds because they are too cool 
(e.g., \citealt{Lamers1999}).

For iPTF13ast, we have no information about the rise time
of the line emission, although it would be very rapid for
a compact progenitor because the star is so small.  
Radiative acceleration does, however, provide a natural
explanation for the broad line wings, obviating the need 
for a fast pre-existing wind.  With the reduced wind speed, the mass
loss rates become more reasonable than the 
$\dot{M} \sim 10^{-2}M_\odot$/year implied by the 
recombination rate and $v_w \sim 10^{3}$~km/s.
Moreover, most models of Type~IIb SNe invoke binary mass 
transfer (e.g., \citealt{Maund2004}, \citealt{Benvenuto2013}, 
\citealt{Bersten2014} but see \citealt{Maund2015}), which can again
provide a high density region to boost the recombination rates
above those for an isolated star with the same mass loss rate.

\section*{Acknowledgments}

CSK thanks S. Johnson, M. Pinsonneault, R. Pogge, A. Piro and T. Thompson for extensive discussions.
CSK is supported by NSF grants AST-1515876 and AST-1515927.  

\appendix
\section{Including Time Delays}
\label{sec:parabaloid}

The effects of light propagation times can be incorporated relatively
easily using the parabolic coordinates $x = \sigma\tau \cos\phi$,
$y= \sigma\tau\sin\phi$ and $z=(\tau^2-\sigma^2)/2$. The radius
is $r = (\sigma^2+\tau^2)/2$, and for an observation at time $t$
the light was emitted at time $t - \sigma^2/c$.  The volume
element is $dV = \sigma \tau(\sigma^2+\tau^2)d\sigma d\tau d\phi$.
The recombination rate of the system compared to that of a
fully ionized wind starting at the stellar surface is
($\Gamma = \Gamma_\infty f(t)$)
\begin{equation}
   f(t) = { R_* \over 4 \pi } \int { dV \over r^4 } g(r,t)
    = 8 R_* \int { \sigma \tau g(r,t) d\sigma d\tau \over (\sigma^2+\tau^2)^3 }
      \label{eqn:parab1}
\end{equation}
where $g(r,t)$ represents any time-dependent recombination.
The limits of integration are from $\sigma_{s}$ to $\sigma_{bo}$
where $0 = 2 R_{s}(\sigma_{s}) -\sigma_{s}^2$ and
$0 = 2 R_{bo}(\sigma_{bo}) -\sigma_{bo}^2$.  For
a given $\sigma$ the limits on $\tau$ are from 
$\tau^2 = \hbox{max}(0, 2 R_{s}(\sigma) - \sigma^2)$  
to $\tau^2 = 2 R_{bo}(\sigma) - \sigma^2$.

If there is no additional function $g(r,t)$, the
$\tau$ integral is analytic in terms of $R_{s}(t)$ 
and $R_{bo}(t)$,
\begin{eqnarray}
   && f(t)={ R_* \over 2} \int_0^{\sigma_{s}}
           \sigma d\sigma \left( { 1 \over R_{s}(\sigma)^2 }
                - { 1 \over R_{bo}(\sigma)^2 }\right) + \nonumber \\
 &&{ R_*\over 2} \int_{\sigma_{s}}^{\sigma_{bo}}
    d\sigma \left( { 4 \over \sigma^3} - {\sigma \over R_{bo}(\sigma)^2 }\right)
      \label{eqn:parab2}
\end{eqnarray}
For inner and outer radii expanding at $v_s$ and
$c$, respectively, only the first integral contributes
for $t < 2R_*/c$ with $\sigma_{s}^2= ct$.  At
later times, 
$\sigma_{s}^2 = 2(R_*+v_s t)/(1+2 v_s/c)$
and
$\sigma_{bo}^2=2(R_*+ct)/3)$.
This provides the analytic solution given in
Equations~\ref{eqn:delay1} and \ref{eqn:delay2}.

\section{HII Regions From Secondaries}
\label{sec:h2region}

Hot main sequence stars may have lower momentum winds than RSGs,
but they also produce ionizing photons which enhance their ability
to perturb the RSG wind by driving the formation of an HII region around
the secondary.  Outside of \cite{Braun2012}, there seems to be no
discussion of this physical effect.  The natural scale for the
ionizing flux from the secondary is ionizing photon production rate
needed to fully ionize the primary's wind,
\begin{equation}
   Q_{crit} = { \dot{M}^2 \alpha_R \over 4 \pi v_w^2 \mu^2 m_p^2 R_*}
      \simeq 1.7 \mu^{-2} \dot{M}^2_4 \alpha_{13} v_{w10}^{-2} R_{*3}^{-1}
        \times 10^{51} \hbox{s}^{-1}.
\end{equation}
This means that an 05 (B0) companion produces enough ionizing photons to
fully photoionize a wind with $\dot{M} \simeq 10^{-5} M_\odot$/year
($10^{-6}M_\odot$/year).
As a toy model, suppose the secondary produces ionizing photons at
rate $Q$ and we assume the density distribution along the line
from the primary to the secondary is unperturbed by the photoionization.
In this case, the radius of the equilibrium photoionization front
from the primary is
\begin{equation}
   { r \over R_*} = \left[ { Q \over Q_{crit}} + { R_* \over D } \right]^{-1}
\end{equation}
where $D$ is again the binary separation.  For comparison, the
stagnation point of the colliding wind shock is at
$R_s = D \beta^{1/2}/(1+\beta^{1/2})$.  In short, a hot main
sequence companion without a wind can likely perturb the wind from
the RSG almost as effectively as one with a fast, low density wind.
In practice, both effects will be present.

{\end{document}